%% file: main.tex
\documentclass[sigconf]{acmart}
\usepackage{url}
\usepackage{array}
\usepackage{graphicx}
\usepackage{booktabs}
\usepackage{amsmath}
\usepackage{wrapfig}
\usepackage{multirow}
\usepackage{subcaption}
\usepackage{fvextra}
\usepackage{listings}
\usepackage{float}

\AtBeginDocument{%
  }

\copyrightyear{2025}
\acmYear{2025}
\setcopyright{cc}
\setcctype{by}
\acmConference[WWW Companion '25]{Companion Proceedings of the ACM Web
Conference 2025}{April 28-May 2, 2025}{Sydney, NSW, Australia}
\acmBooktitle{Companion Proceedings of the ACM Web Conference 2025 (WWW
Companion '25), April 28-May 2, 2025, Sydney, NSW, Australia}
\acmDOI{10.1145/3701716.3715225}
\acmISBN{979-8-4007-1331-6/2025/04}

\begin{document}

\title{Flow-of-Action: SOP Enhanced LLM-Based Multi-Agent System for Root Cause Analysis}

\author{Changhua Pei}
\email{chpei@cnic.cn}
\affiliation{%
  \institution{Computer Network Information Center, Chinese Academy of Sciences}
  \city{Beijing}
  \country{China}
}
\additionalaffiliation{
  \institution{Hangzhou Institute for Advanced Study, University of Chinese Academy of Sciences}
  \city{Hangzhou}
  \country{China}
}

\author{Zexin Wang}
\email{wangzexin@cnic.cn}
\affiliation{%
  \institution{Computer Network Information Center, Chinese Academy of Sciences}
  \city{Beijing}
  \country{China}
}
\additionalaffiliation{%
  \institution{University of Chinese Academy of Sciences}
  \city{Beijing}
  \country{China}
}
\authornote{Also with ByteDance. Work done during the internship at ByteDance.}

\author{Fengrui Liu}
\email{liufengrui.work@bytedance.com}
\author{Zeyan Li}
\email{lizeyan.42@bytedance.com}
\affiliation{%
  \institution{ByteDance}
  \city{Beijing}
  \country{China}
}

\author{Yang Liu}
\authornotemark[2]
\email{liuyang@cnic.cn}
\affiliation{%
  \institution{Computer Network Information Center, Chinese Academy of Sciences}
  \city{Beijing}
  \country{China}
}

\author{Xiao He}
\email{xiao.hx@bytedance.com}
\affiliation{%
  \institution{ByteDance}
  \city{Hangzhou}
  \country{China}
}

\author{Rong Kang}
\email{kangrong.cn@bytedance.com}
\affiliation{%
  \institution{ByteDance}
  \city{Beijing}
  \country{China}
}

\author{Tieying Zhang}
\authornotemark[4]
\email{tieying.zhang@bytedance.com}
\author{Jianjun Chen}
\email{jianjun.chen@bytedance.com}
\affiliation{%
  \institution{ByteDance}
  \city{San Jose}
  \country{United States}
}

\author{Jianhui Li}
\authornote{ Corresponding Authors.}
\email{lijh@cnic.cn}
\author{Gaogang Xie}
\email{xie@cnic.cn}
\affiliation{%
  \institution{Computer Network Information Center, Chinese Academy of Sciences}
  \city{Beijing}
  \country{China}
}

\author{Dan Pei}
\email{peidan@tsinghua.edu.cn}
\affiliation{%
  \institution{Tsinghua University}
  \city{Beijing}
  \country{China}
}

\renewcommand{\shortauthors}{Changhua Pei et al.}

\begin{abstract}
In the realm of microservices architecture, the occurrence of frequent incidents necessitates the employment of Root Cause Analysis (RCA) for swift issue resolution. It is common that a serious incident can take several domain experts hours to identify the root cause. Consequently, a contemporary trend involves harnessing Large Language Models (LLMs) as automated agents for RCA. Though the recent ReAct framework aligns well with the Site Reliability Engineers (SREs) for its thought-action-observation paradigm, its hallucinations often lead to irrelevant actions and directly affect subsequent results. Additionally, the complex and variable clues of the incident can overwhelm the model one step further. To confront these challenges, we propose \textbf{Flow-of-Action}, a pioneering Standard Operation Procedure (SOP) enhanced LLM-based multi-agent system. By explicitly summarizing the diagnosis steps of SREs, SOP imposes constraints on LLMs at crucial junctures, guiding the RCA process towards the correct trajectory. To facilitate the rational and effective utilization of SOPs, we design an SOP-centric framework called \textbf{SOP flow}. SOP flow contains a series of tools, including one for finding relevant SOPs for incidents, another for automatically generating SOPs for incidents without relevant ones, and a tool for converting SOPs into code. This significantly alleviates the hallucination issues of ReAct in RCA tasks. We also design multiple auxiliary agents to assist the main agent by removing useless noise, narrowing the search space, and informing the main agent whether the RCA procedure can stop. Compared to the ReAct method's 35.50\% accuracy, our Flow-of-Action method achieves 64.01\%, meeting the accuracy requirements for RCA in real-world systems.
\end{abstract}

\ccsdesc[300]{Software and its engineering~Software maintenance tools}

\keywords{Root Cause Analysis, Multi-Agent System, Large Language Model}

\maketitle

\section{Introduction}
\input{intro}
\label{intro}

\section{Flow-of-Action}
\input{method}
\label{method}

\section{Evaluation}
\input{evaluation}
\label{evaluation}

\section{Conclusion}
\input{conclusion}
\label{conclusion}

\section{Acknowledgment}
\input{acknowledgement}


\bibliographystyle{ACM-Reference-Format}
\bibliography{ref}

\appendix

\section{Multimodal Data Collection}
\label{multidata}

\begin{figure}[H]
\centerline{\includegraphics[scale=0.24]{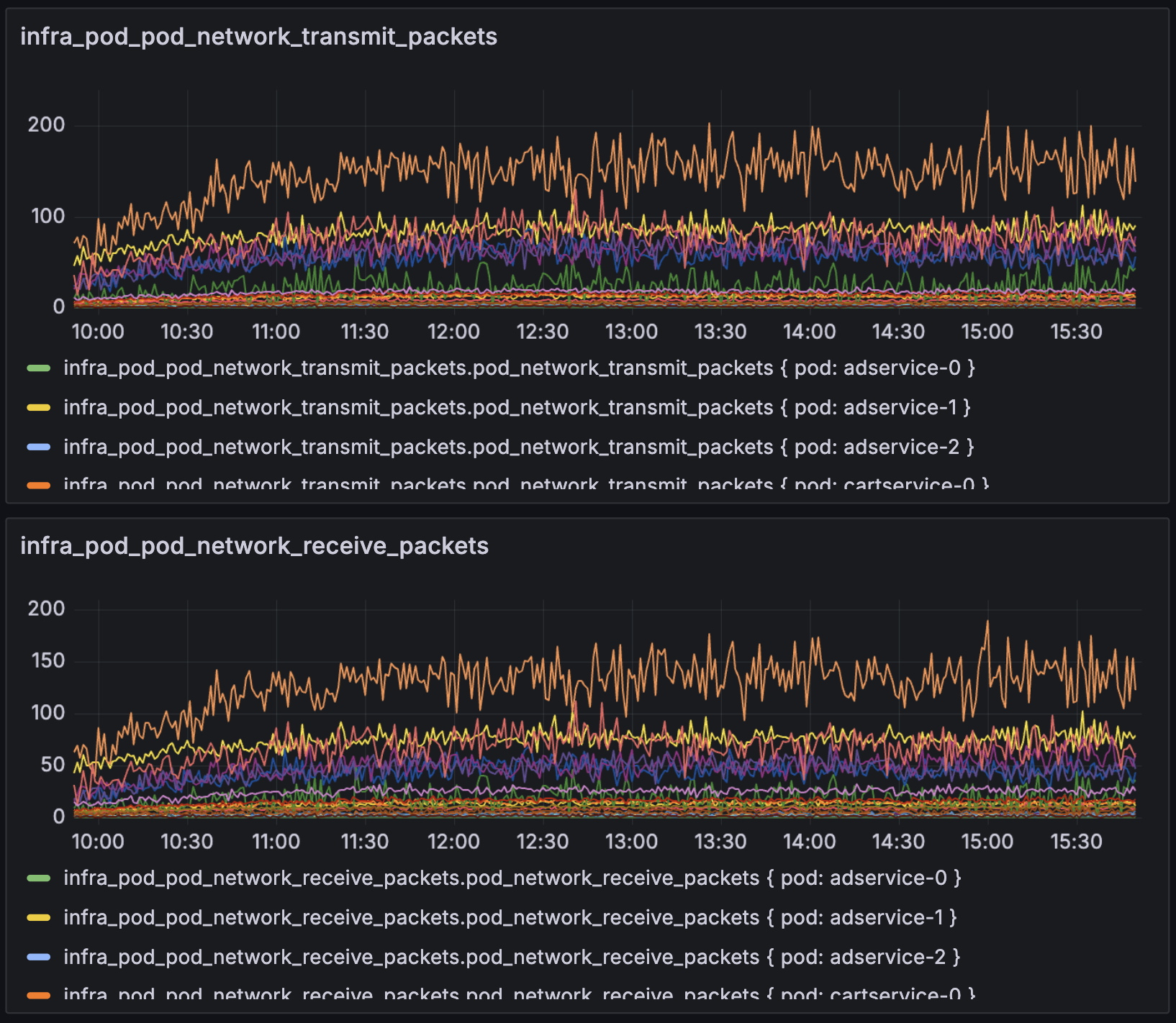}}
\caption{Prometheus Dashboard.}
\label{prometheus}
\end{figure}

\begin{figure}[H]
\centerline{\includegraphics[scale=0.14]{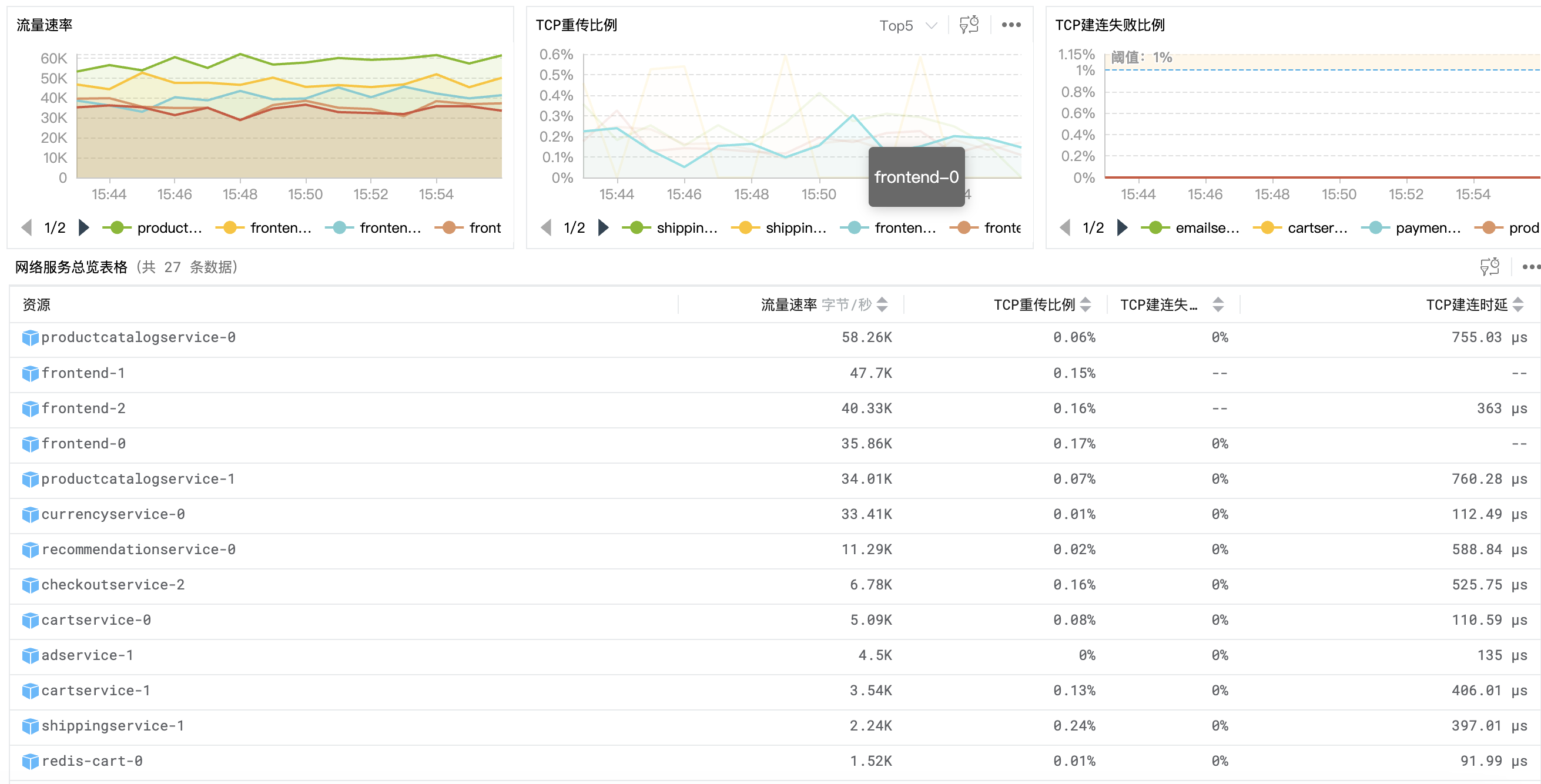}}
\caption{Deepflow Dashboard.}
\label{deepflow}
\end{figure}

\begin{figure}[H]
\centerline{\includegraphics[scale=0.12]{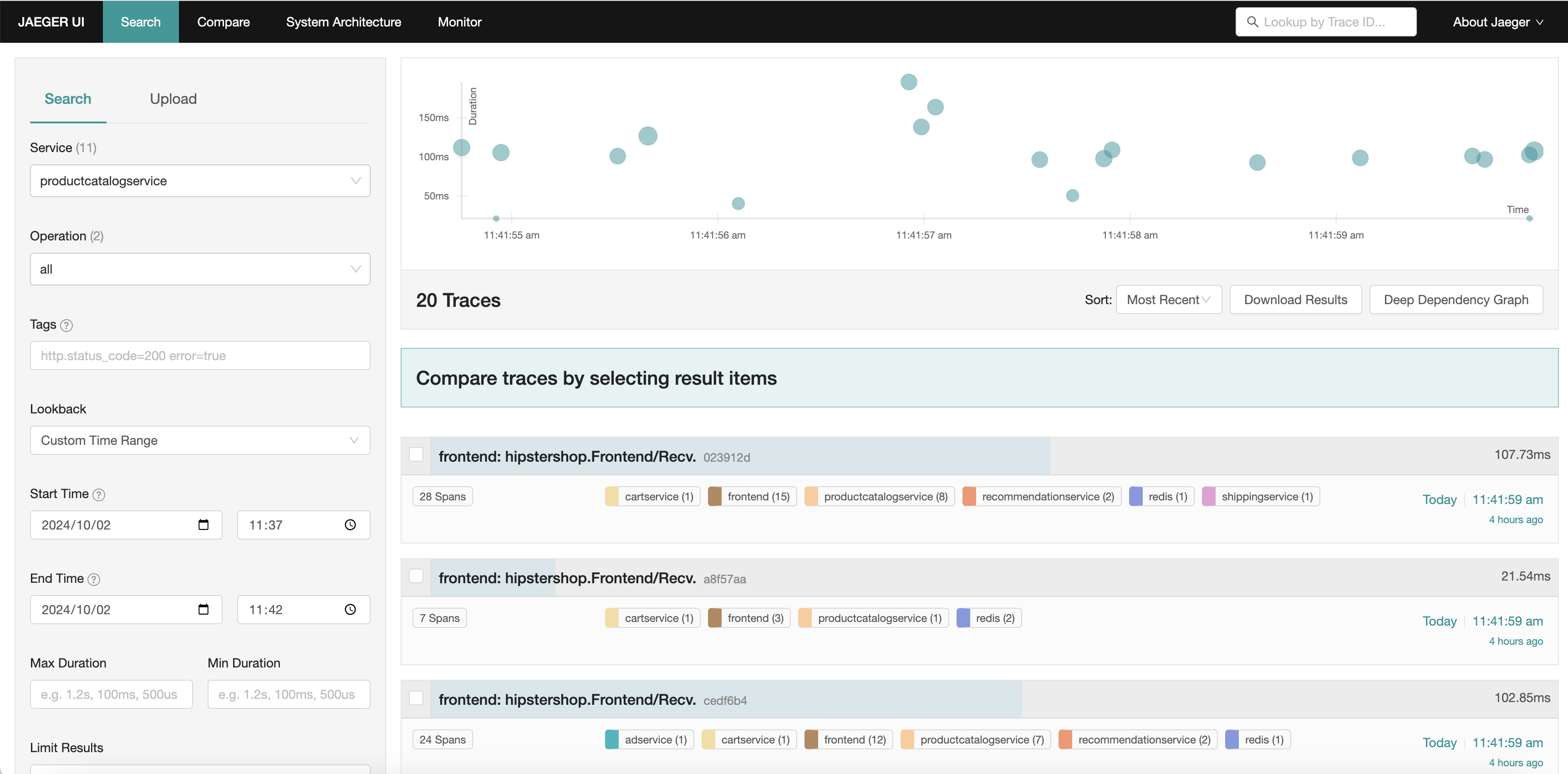}}
\caption{Jaeger Dashboard.}
\label{jaejer}
\end{figure}

\begin{figure}[H]
\centerline{\includegraphics[scale=0.12]{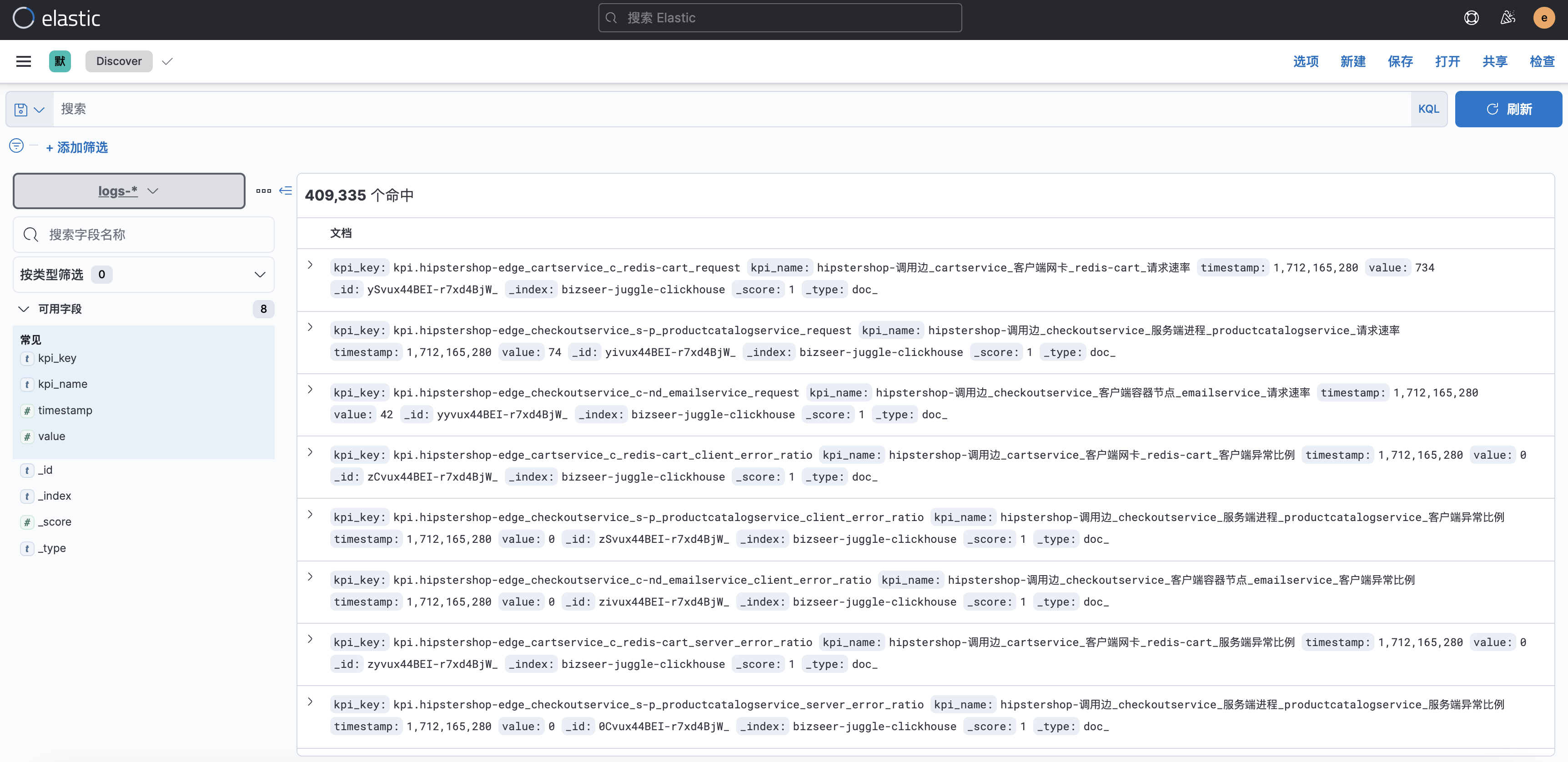}}
\caption{Elastic Dashboard.}
\label{elastic}
\end{figure}

We first deploy various data collection systems (Figure \ref{prometheus}, Figure \ref{deepflow}, Figure \ref{jaejer}, Figure \ref{elastic}). For metrics, we start by deploying Prometheus, which collects architecture-level metrics, such as pod-level and node-level indicators that are generally standardized and unrelated to business logic. Additionally, we deploy DeepFlow to gather business-level metrics, such as business traffic data. For anomaly detection, we use traditional rule-based methods because they are fast and convenient.

For trace data, we deploy Jaeger to collect all trace data, where each trace represents a call chain containing multiple spans, with each span corresponding to a single call. Anomalies can occur within any span. In the current environment, detecting trace anomalies is relatively straightforward, as a span failure typically includes an associated error message. Therefore, we directly extract error messages to generate alert reports. For log data, we use Elastic for collection. Since abnormal logs usually contain specific keywords, extracting anomalies based on keywords has become widely accepted. We also adopt this keyword-based approach for log anomaly detection.


\section{Tools}
\label{apendixb}

\begin{table}[H]
\centering
\caption{Description of Tools}
\label{tab:tools}
\small
\begin{tabular}{p{0.15\textwidth}p{0.22\textwidth}} 
\toprule
\multicolumn{1}{c}{Tool} & \multicolumn{1}{c}{Description}                                                                                                                                                                                                                                                                                                     \\ 
\midrule
pod\_analyze             & Analyzing all pods' status.                                                                                                                                                                          \\
node\_analyze            & Analyzing all nodes' status.                                                                                                                                                                     \\
service\_analyze         & Analyzing all services' status.                                                                                                                                                         \\
deployment\_analyze      & Analyzing all deployments' status.    \\
statefulset\_analyze     & Analyzing all statefulsets' status.                                                                                                                                                                                       \\
run\_kubectl\_command    & Executing kubectl commands generated by LLMs.                                                                                                                                          \\
get\_all\_namespace      & Obtaining a list of all namespaces.                        \\

get\_relevant\_metric      & Obtaining relevant metric names according to query.                        \\
\bottomrule
\end{tabular}
\end{table}

\end{document}

%% file: intro.tex
\begin{figure*}[htbp]
\centerline{\includegraphics[scale=0.3]{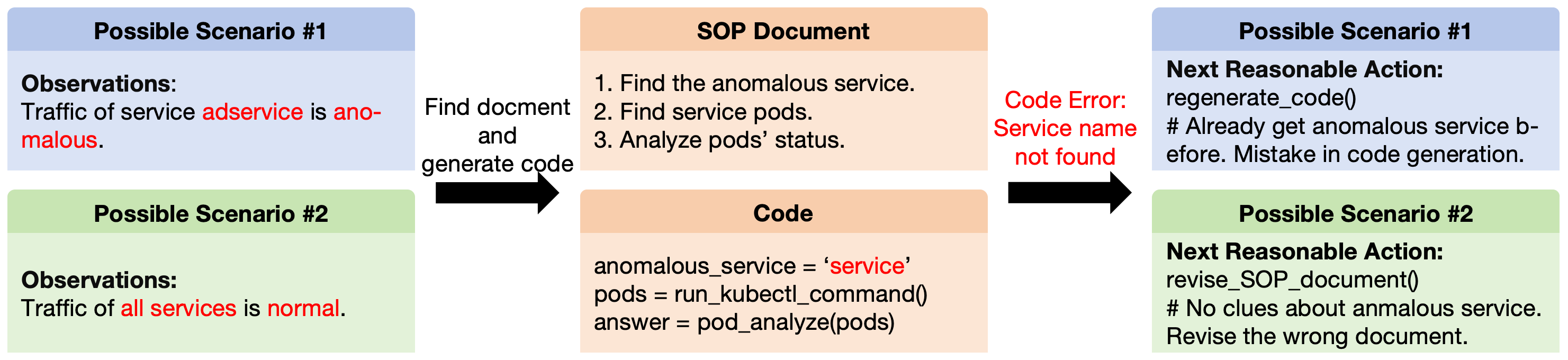}}
\caption{Illustration example of challenge 2.}
\label{challenge2}
\end{figure*}

In today's large-scale web systems and services, traditional monolithic applications encounter notable challenges including intricate deployment processes and limited scalability \cite{micro1,micro2,micro3,micro4,micro5}, attributed to the proliferation of services and frequent service iterations. In response to this context, Microservices Architecture (MSA) has surfaced and continually evolved \citep{chen2024microfi}. By disassembling monolithic applications into small, self-sufficient service units, each dedicated to specific business functionalities, MSA presents benefits such as loose coupling, independent deployment, and effortless scalability. Nevertheless, with the escalation of user numbers and their corresponding demands, the diversity and quantity of MSA instances also increase. Despite the implementation of numerous monitor tools, recurrent incidents arise from hardware malfunctions or misconfigurations, posing challenges to reliability assurance. These incidents lead to substantial financial losses \cite{wang2024large}. For instance, on November 12, 2023, Alibaba experienced a large-scale outage, resulting in the interruption of multiple services for nearly three hours\footnote{https://www.datacenterdynamics.com/en/news/alibaba-cloud-hit-by-outage-second-in-a-month/}.


To promptly tackle these incidents, Root Cause Analysis (RCA) has emerged as a prominent research area within Artificial Intelligence for IT Operations (AIOps) in recent years \cite{rca1,rca2,rca3,rca4}. Traditional RCA techniques, in order to address the difficulties of manual fault diagnosis, have employed deep learning methods to learn from historical faults \citep{li2022dejavu}. However, these methods have two main drawbacks. First, they have poor adaptability to new scenarios, requiring model retraining when faced with a new situation. Second, they only output the root cause of the fault without providing the entire diagnostic process, resulting in poor explainability. This situation often results in Site Reliability Engineers (SREs) harboring a sense of distrust towards the results, as they fear that misidentifying the root cause could potentially result in further wasted repair time or exacerbate faults by addressing the wrong issue. Over the recent years, Large Language Model (LLM) agents like ReAct \citep{yao2022react} and ToolFormer \citep{schick2024toolformer} have been deployed across diverse domains. LLM agents harness their robust natural language understanding capabilities to adeptly coordinate various tools, allowing SREs to see the entire troubleshooting process and providing rich explanations for the root causes. Nonetheless, despite the considerable prowess of LLM agents, the efficient and accurate utilization of LLM agents in RCA encounters ongoing challenges.

\noindent\textbf{Challenge 1: Randomness and hallucinations leading to irrational action selection}


Current LLMs primarily function as probabilistic models \citep{radford2018gpt,radford2019gpt2}, thereby exhibiting pronounced randomness and tendencies towards generating hallucinations. Employing an LLM agent for RCA activities necessitates the retrieval and comprehension of diverse data modalities (metric \citep{metric6}, log \citep{log2}, trace \citep{yaosparserca}) and the extensive utilization of API tools. As the scope of the context expands, issues often emerge such as inaccurate parameter extraction leading to failures in tool invocation and discrepancies between tool invocations and the context at hand. Instances of randomness or hallucinations at any stage can significantly impact the subsequent trajectory of the RCA procedure, hindering the accurate identification of the true root cause.

\noindent\textbf{Challenge 2: Complex and variable observations leading to multiple reasonable actions}



Existing LLM agents are typically bundled with a diverse array of tools \citep{qin2023toolllm}, especially within complex domains like RCA, where the number of APIs can escalate to hundreds. Each API invocation results in varied observations, thereby introducing intricacies in action selection. Furthermore, even when confronted with identical observations, multiple plausible actions may be viable. For example, as shown in Figure \ref{challenge2}, within the context of a code error ``Service name not found'', the root cause could originate from errors in the code generation phase or inaccuracies in associated SOP document, prompting multiple feasible actions like code regeneration or document revision.

\begin{figure*}[htbp]
\centerline{\includegraphics[scale=0.3]{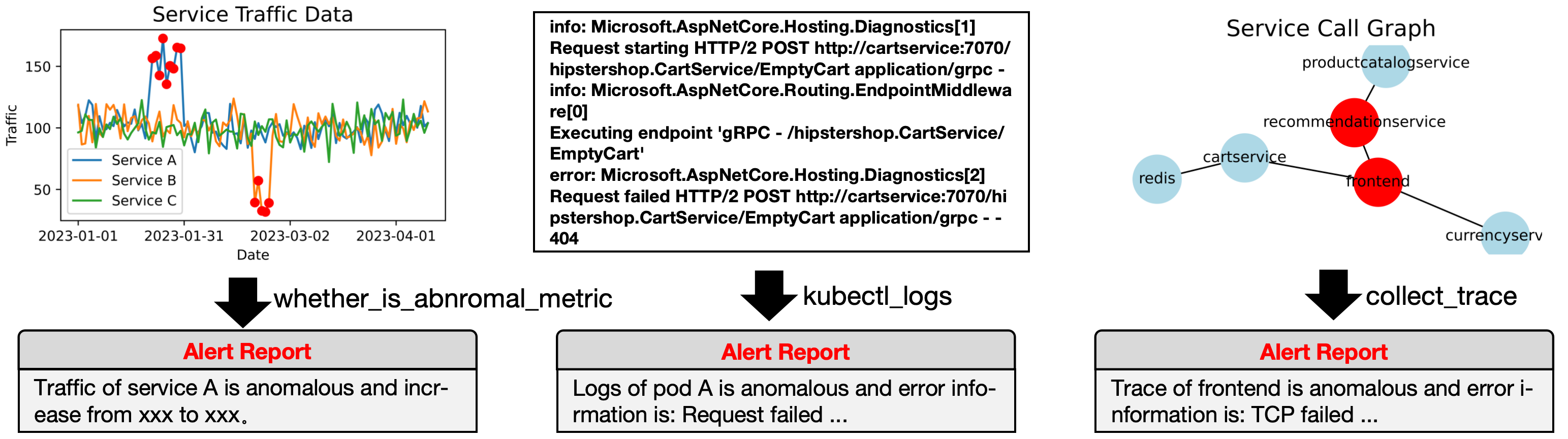}}
\caption{Multimodal data collection and analysis.}
\label{fig:multimodal}
\end{figure*}

To confront the challenges outlined above, we propose \textbf{Flow-of-Action}, a Standard Operating Procedure (SOP) enhanced Multi-Agent System (MAS). Initially, to mitigate the impact of randomness and hallucinations in the orchestration process, we integrate SOPs into the knowledge base and propose the \textbf{SOP flow}. Specifically, SOPs outline a standardized set of steps for RCA, while SOP flow represents an efficient and accurate process built upon SOPs for their effective utilization. Through prompt engineering, we ensure that the orchestration of the main agent loosely follows the SOP flow in the absence of unexpected circumstances. Subsequently, to tackle the second challenge, compared with the thought-action-observation paradigm, we propose the thought-\textbf{actionset}-action-observation paradigm. Flow-of-Action avoids immediate action selection and instead generates a reasoned action set before making the final decision on the course of action. Besides, we devise a novel MAS. Specifically, we introduce multiple agents such as MainAgent, CodeAgent, JudgeAgent, ObAgent, and ActionAgent, each entrusted with distinct responsibilities, collaborating harmoniously to enhance root cause identification. 

Our key contributions are summarized as follows:
\begin{itemize}
    \item We propose the Flow-of-Action framework, the first agent-based fault localization process centered around SOPs. With this framework, we significantly reduce the inefficiency in action selection of the native ReAct framework and reducing the cost of trial and error.

    \item We introduce the concept of SOPs to integrate the expert experience into the LLM to greatly reduce hallucinations during RCA. For any given fault, we can automatically match the most relevant set of SOPs and can also generate new SOPs automatically, extending the limited set of human-generated SOPs.

    \item We innovatively propose a multi-agent collaborative system, including JudgeAgent and ObAgent. JudgeAgent assists the MainAgent in determining whether the root cause of the fault has been identified in the current iteration, while ObAgent helps MainAgent extract fault types and key information from massive amounts of data, addressing the information overload issue in the RCA process.

    \item Through a fault-injection simulation platform of a real-world e-commerce system, Flow-of-Action has increased the localization accuracy from 35\% to 64\% compared to ReAct, proving the effectiveness of the Flow-of-Action framework.

\end{itemize}

%% file: method.tex

    

\begin{figure*}[htbp]
\centerline{\includegraphics[scale=0.32]{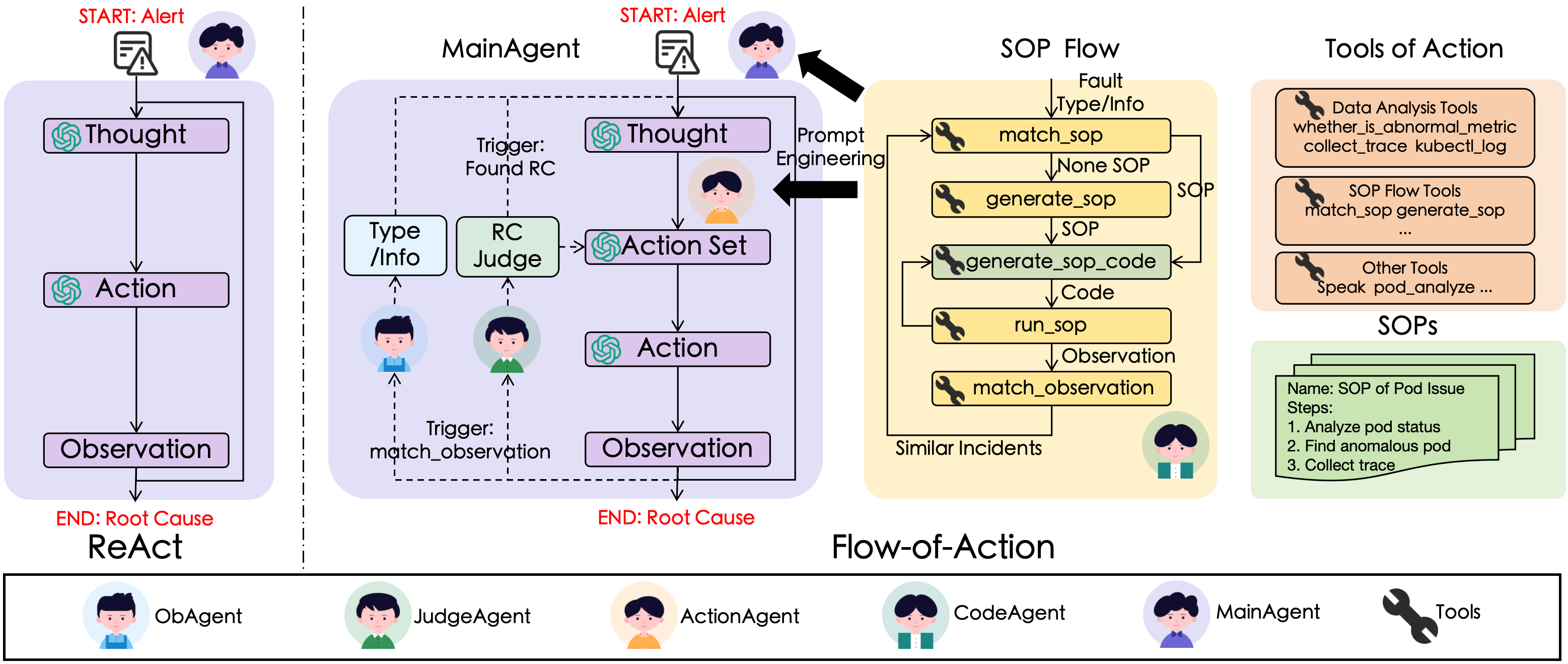}}
\caption{Comparison of ReAct and Flow-of-Action. RC means root cause. Dashed lines represent paths triggered under specific conditions. When the previous action is $match\_observation$, JudgeAgent and ObAgent are triggered. When JudgeAgent finds the root cause, it triggers the input of the analysis result to thought and adds $Speak$ to action set.}
\label{fig:model}
\end{figure*}

In this section, we will present the design of Flow-of-Action. As illustrated in Figure \ref{fig:model}, the Flow-of-Action is a MAS built upon the ReAct. It encompasses three key design components: the SOP flow, the action set, and the MAS. We will delve into each of these components in the subsequent sections. Prior to their detailed exploration, we will introduce the foundational knowledge required, including the knowledge base and tools utilized by the Flow-of-Action.

\subsection{Knowledge Base of Agents}

Given the restricted context length of LLMs, Retrieval-Augmented Generation (RAG) has experienced notable progress \citep{jeong2024adaptive}. However, the quality of text retrieved by RAG significantly influences the ultimate outcomes. Many existing RAG methodologies segment documents within the knowledge base and employ semantic block embeddings to calculate similarity for retrieval. This approach, however, does not consistently yield optimal results in RCA. Therefore, we have devised an innovative knowledge base model integrating SOP knowledge and historical incident knowledge. 

    

\subsubsection{SOP Knowledge}


With the successful integration of SOPs in the realm of code generation \citep{hong2023metagpt}, there is a growing recognition that relying solely on LLMs to execute intricate tasks like RCA is impractical. SOPs, to a certain extent, impose constraints on LLMs at crucial junctures, guiding the entire process towards the correct trajectory. Consequently, we have embedded SOPs into the knowledge base, which are either authored by engineers based on domain expertise or extracted through automation tools. As shown in Figure \ref{fig:model}, each SOP constitutes a self-contained unit comprising two attributes: name and steps. The name encapsulates essential information about the SOP, which is translated into a vector for subsequent retrieval purposes.

\subsubsection{Historical Incidents}

As highlighted by \citet{chen2024automatic}, in systems where similar incidents occur frequently, historical incident data proves invaluable in identifying the root cause of ongoing incidents. Consequently, we incorporate the performance details of historical incidents into the knowledge base. Each historical incident is characterized by two key attributes: manifestation and type. When retrieving similar incidents, we evaluate similarity by comparing the embedding of the current observation with the embedding of the manifestation of historical incidents. However, relying solely on embeddings for assessment can introduce significant errors. To tackle this issue, we have intentionally devised the ObAgent (elaborated upon subsequently) to address this challenge.

\subsection{Tools of Agents}

Within LLM agents, tools typically refer to pre-defined functions. During the action phase, LLM invokes relevant tools to obtain the necessary information. In Flow-of-Action, the tools utilized primarily fall into three categories: tools for multimodal data collection and analysis, tools related to SOP flow, and other tools. Each category will be discussed in detail below.


\subsubsection{Multimodal Data Collection and Analysis}


\begin{table*}
\centering
\caption{Description of SOP flow tools.}
\label{tab:frameworktool}
\small
\begin{tabular}{lccc} 
\toprule
                    & Input                   & Output                        & LLM Usage  \\ 
\midrule
match\_sop          & Fault Type/Information & SOP                           & No         \\
generate\_sop       & Fault Type/Information & SOP                           & Yes        \\
generate\_sop\_code & SOP                     & SOP Code                      & Yes        \\
run\_sop            & SOP Code                & Result after running the code & No         \\
match\_observation  & Observation             & Similar incidents             & No         \\
\bottomrule
\end{tabular}
\end{table*}

Within the realm of MSA, which encompasses diverse modalities of data such as metrics, traces, and logs, the importance of multimodal data for RCA has been underscored by existing methodologies \citep{yao2024chain,yu2023nezha}. Consequently, we have implemented a comprehensive monitoring system to aggregate multimodal data. While LLMs excel in processing textual data, their effectiveness in interpreting structured data types like metrics is constrained, especially in the presence of data noise. Therefore, it is imperative to preprocess the data by denoising and transforming it into textual format for enhanced comprehension by LLMs. As depicted in Figure \ref{fig:multimodal}, we have devised the following components: $whether\_is\_abnormal\_metric$ to leverage time series anomaly detection algorithms \citep{wang2024revisiting,donut,tuli2022tranad} for identifying metric anomalies and converting them into fault-related text; $collect\_trace$ for capturing abnormal span details across the entire call chain and converting them into text format; and $kubectl\_logs$ for extracting abnormal log information from each pod within the Kubernetes system.

\subsubsection{SOP Flow Tools}


As previously mentioned, we have introduced a flow centered around SOPs. This comprehensive flow is meticulously crafted based on common workflows employed by SREs in practical settings, integrating innovative concepts such as code. Details regarding the tools utilized within the flow are delineated in Table \ref{tab:frameworktool}. Moreover, to preempt unexpected incidents during the flow's operation, we have developed a variety of targeted auxiliary tools. For example, within the context of $generate\_sop$, we have introduced $get\_relevant\_metric$ to streamline the retrieval of pertinent metric names.

\subsubsection{Other Tools}


The flow aims to establish a standardized and generalized process for intricate RCA tasks, devoid of service- or business-specific components within the tools themselves. However, a broader array of tools is necessitated when generating SOPs or SOP code, or when executing operations beyond the flow, to query the authentic operational state of the system. In addition to the previously mentioned tools for querying and analyzing multimodal data, a suite of tailored analysis tools has been devised for MSA, including $pod\_analyze$ and $service\_analyze$. These tools employ queries on specific attribute data within the Kubernetes system to ascertain the system's status. Upon identification, $Speak$ is employed to communicate the discovered root cause to all pertinent stakeholders. For a comprehensive elucidation of these tools, kindly consult the Appendix \ref{apendixb}.

\subsection{SOP Flow}


\begin{figure*}[htbp]
\centerline{\includegraphics[scale=0.23]{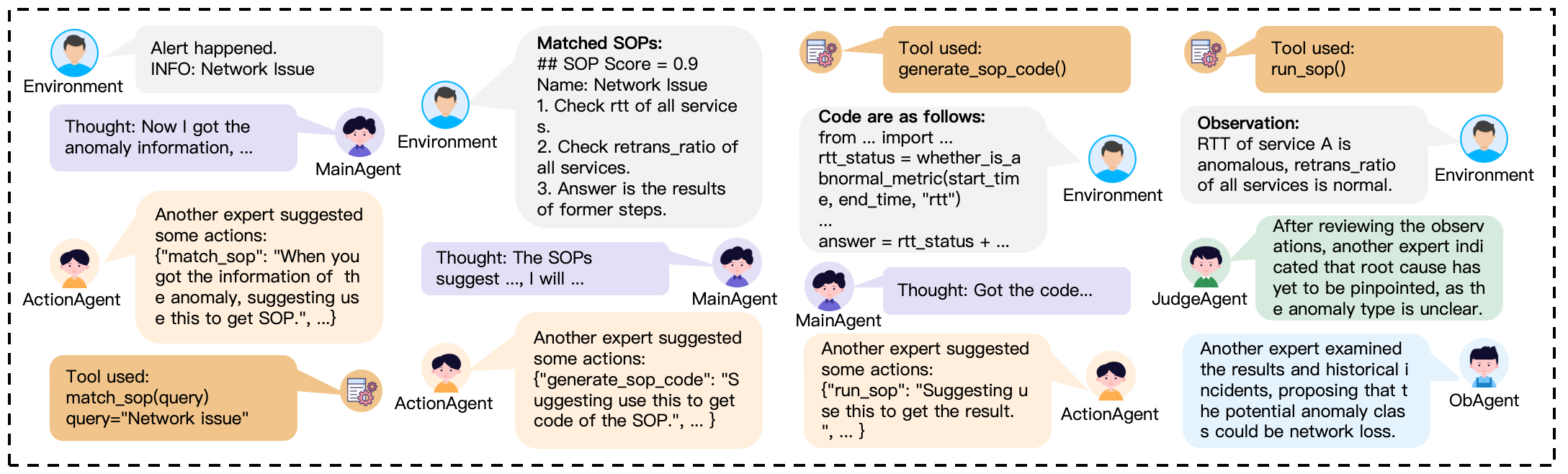}}
\caption{Example of Flow-of-Action.}
\label{fig:example}
\end{figure*}

\begin{figure}[htbp]
\centerline{\includegraphics[scale=0.37]{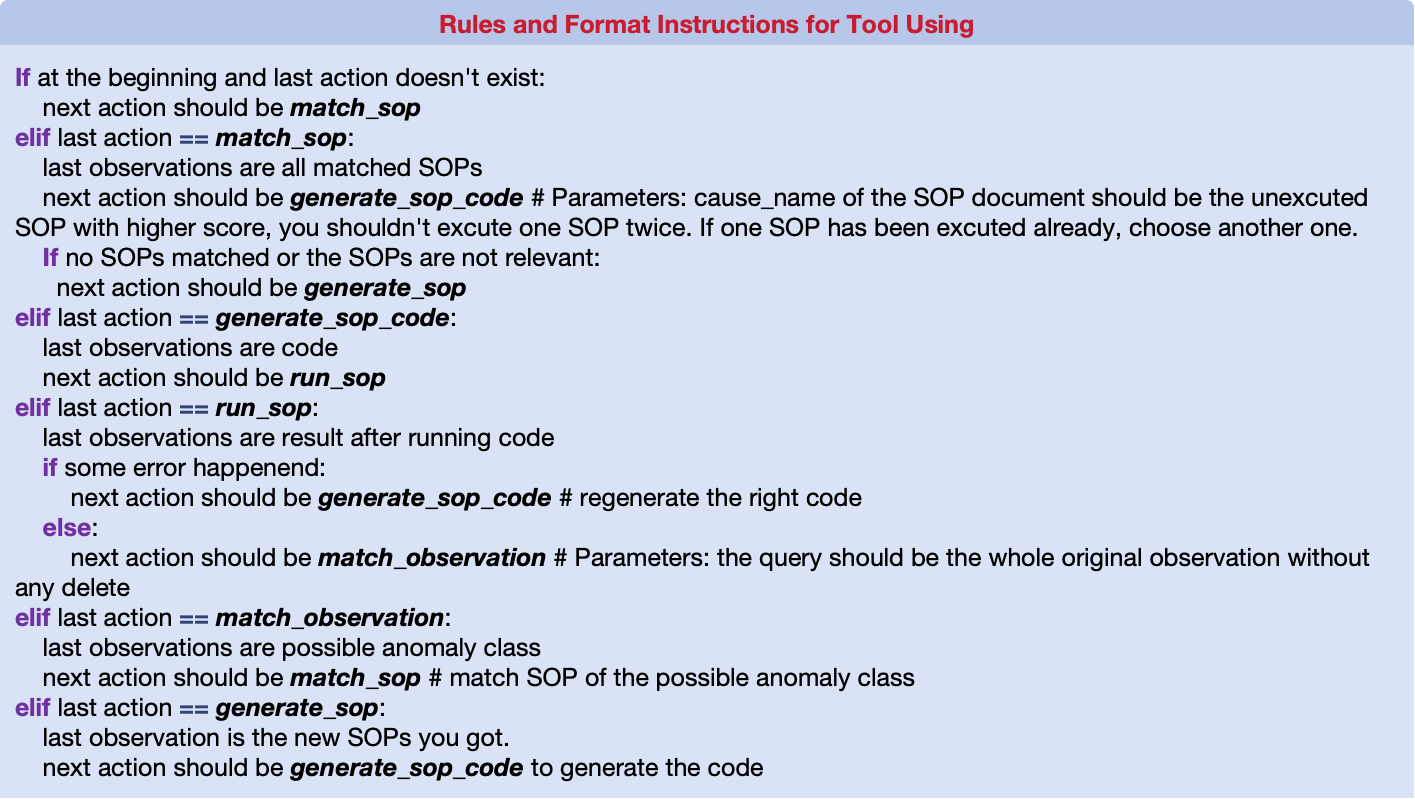}}
\caption{SOP flow prompt information of LLMs.}
\label{fig:promptengineering}
\end{figure}

The SOP flow represents a comprehensive logic chain of actions tailored to the SOP mentioned earlier. It serves to instruct LLMs on how to effectively utilize SOP knowledge. For instance, in the initial stages of RCA, it is essential to identify which SOPs are most relevant to the incident (corresponding to $match\_sop$). Additionally, if a particular incident does not align with any existing SOP, the automation of SOP generation should be considered (corresponding to $generate\_sop$). While the comprehensive SOP flow can be visually represented, as illustrated in Figure \ref{fig:model}, in practical application, the full SOP flow is presented in the form of prompts to the MainAgent to aid in thought processes and to the ActionAgent to generate a more rational action set. The SOP flow prompt information provided to the LLMs is displayed in Figure \ref{fig:promptengineering}. By implementing such soft constraints, we aim to tackle the issue of chaotic tool orchestration while still maintaining the flexibility of LLMs. Unlike methods like FastGPT \citep{fastgpt}, we do not enforce strict workflow constraints on LLM orchestration. Figure \ref{fig:example} provides an example of the Flow-of-Action. Subsequently, we will systematically elucidate critical transitional subflows within the SOP flow.

\subsubsection{Fault Type/Information$\rightarrow$SOP}

\begin{figure}[htbp]
\centerline{\includegraphics[scale=0.6]{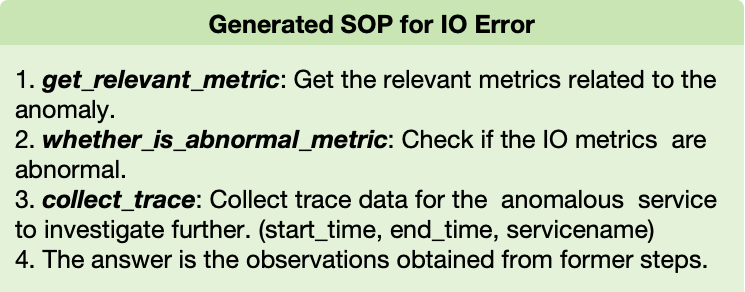}}
\caption{Generated SOP for IO error.}
\label{fig:ioerror}
\end{figure}

In our flow, we initially utilize $match\_sop$ to associate the fault information with the relevant SOP. This matching process involves computing the similarity between the current query and all SOP name embeddings, ranking them, and selecting the top \(k\) matches. To avoid matching with highly irrelevant SOPs, a filtering threshold is established. Nevertheless, in real-world contexts where new fault types frequently emerge, instances may arise where pertinent SOPs cannot be matched. To tackle this challenge, we introduce $generate\_sop$ to devise new SOPs for queries that do not align with existing SOPs. Specifically, we utilize LLMs to generate new SOPs and leverage existing SOPs as few-shot prompts to guide the development of more standardized and coherent SOPs. Figure \ref{fig:ioerror} shows an example of an SOP for handling IO errors generated by an LLM. Although not highly precise, the general direction of analysis is correct. The logic is rigorous, and the diagnostic results provide useful assistance to SREs.

Within the entirety of the flow, the generation of SOPs stands as a pivotal phase as it directly influences the subsequent RCA process. To enhance the precision of RCA, we have devised hierarchical SOPs. Our objective is for the RCA process to progress from a macro to micro level, from a general to specific perspective, mirroring real-world scenarios more closely. For instance, we first address network issues before delving into network partition problems.

\subsubsection{SOP$\rightarrow$SOP Code}

Once a suitable SOP is obtained, due to the interdependence of steps within the SOP, it is generally necessary to execute the SOP step by step to achieve the desired outcome. However, in real-world scenarios, SOPs are typically concise texts, making it relatively difficult for engineers lacking domain knowledge to execute the entire SOP. Utilizing an agent based on LLM to execute the SOP is a more rational and efficient approach. However, directly instructing the agent to execute all steps of the SOP one by one often leads to errors. This is because LLM tends to focus more on proximal text, and the outcome of a particular step can significantly influence the selection of subsequent actions.

Therefore, we have designed $generate\_sop\_code$ to convert the entire SOP into code for simultaneous execution. This approach offers three main advantages. Firstly, numerous works, including Chain-of-Code \citep{li2023chainofcode}, have demonstrated that executing code in LLM environments is far more accurate than executing text \citep{pan2023logiclm}, aligning well with the precise requirements of RCA. Secondly, in many scenarios, including RCA, there exist numerous atomic operations where we wish for several actions to be executed together or none at all, as executing a single action in isolation may not yield useful results. SOPs exemplify this situation, where executing only a portion may not yield the desired fault information. Converting SOPs to code effectively addresses this issue, as once the code is executed, it must run from start to finish. Lastly, SOP code represents a collection of multiple actions, enabling the execution of multiple actions with a single tool invocation, thereby significantly reducing LLM token and resource consumption.

\subsubsection{SOP Code$\rightarrow$Observation}

After obtaining the SOP code, the flow invokes $run\_sop$ to execute the entire SOP code. However, the generation of code is not always accurate and may lead to various issues, such as syntax errors or incorrect variables within the code. In such instances, our flow expects to re-match suitable parameters and use $generate\_sop\_code$ to generate new, correct code. Once the code is error-free, we can smoothly execute it to obtain the desired results.

\subsubsection{SOP Code$\rightarrow$Fault Type/Information}

As mentioned earlier, the definition of SOP is hierarchical, and our RCA process follows a layered and progressive approach. Upon executing $run\_sop$ and obtaining a new observation, we seek guidance to determine the next steps in the localization process. The ideal approach is to identify potential fault types based on the observation. Relying solely on the domain knowledge of the LLM agent is evidently insufficient for accurate judgment in a specific domain, necessitating fine-tuning of the LLM model or the introduction of more domain-specific knowledge. Inspiration from various methods \citep{chen2024automatic} suggests that most fault types have occurred historically. Therefore, we use $match\_observation$ to recall similar historical incidents based on observation. The ObAgent is then utilized to determine potential fault types or provide descriptions of faults for subsequent RCA processes.

\subsection{Action Set}

In section \ref{intro}, we mentioned that in RCA, it is relatively challenging for the LLM agent to perform reasonable planning. This difficulty primarily arises from two reasons: the variability of observations and the existence of multiple possible actions for a given observation. Instantaneously identifying and executing the most reasonable action from numerous viable choices is an exceedingly challenging task for the LLM.

To address this challenge, we have devised a mechanism known as the action set. Specifically, drawing inspiration from the CoT \citep{wei2022chain}, we first generate a series of reasonable actions comprising a set, with each action accompanied by a textual explanation of the rationale behind its selection. This set primarily consists of two components: actions generated by the ActionAgent and actions identified by the JudgeAgent. The ActionAgent incorporates flow information and numerous examples in the prompt to enhance the rationality of the generated actions. However, this may still overlook reasonable flow actions. Therefore, we have established a rule based on the flow to ensure that the action set is comprehensive and logical. For instance, if the preceding action was $generate\_sop$, the subsequent action of $generate\_sop\_code$ is added to the set. Secondly, the JudgeAgent evaluates whether the root cause has been identified during the current RCA process. If the root cause is pinpointed, the action $Speak$ is included in the action set.

Through action set, we have effectively mitigated the challenges posed by diverse observations and a plethora of feasible actions that could potentially hinder agent planning. Furthermore, the strategic design of the action set has enabled the LLM Agent to attain a nuanced equilibrium between stochasticity and determinism. Within RCA, excessive randomness may induce divergence in the localization process, impeding the formation of effective diagnostics. Conversely, an overly deterministic approach may incline the model towards scripted operations, limiting its capacity to handle unforeseeable and rapidly changing circumstances.

\subsection{Multi-Agent System}

We have designed a MAS consisting of a single main agent along with multiple auxiliary agents. The MainAgent serves as the principal entity with authority, while the other agents are responsible for providing suggestions to it. The MainAgent orchestrates the entire localization process. The ActionAgent provides a feasible set of actions for the MainAgent to choose from. The ObAgent offers potential anomaly types or information after the MainAgent completes $match\_observation$. The JudgeAgent determines whether the root cause has been identified. However, even if the JudgeAgent believes the root cause has been found, the MainAgent may not necessarily use $Speak$ to conclude the entire localization process. Taking additional steps and gathering more information may lead to a more accurate root cause determination. The CodeAgent plays a crucial role in the SOP flow, possessing information on all tools and generating appropriate code for subsequent use. Through the MAS, the burden on the MainAgent is significantly reduced. It only needs to consider the opinions of other agents and make relatively accurate judgments based on the entire localization process. Such division of labor also aligns more closely with real-world operational scenarios.

%% file: evaluation.tex



\subsection{Experiment Setup}

\subsubsection{Dataset}
We have deployed the widely used microservices system GoogleOnlineBoutique\footnote{https://github.com/GoogleCloudPlatform/microservices-demo}, an e-commerce system consisting of over 10 services, on the Kubernetes platform. Building upon this, we have implemented Prometheus, Elastic, DeepFlow \cite{deepflow}, and Jaeger to collect metric, log, and trace data (Detailed in Appendix \ref{multidata}). Anomalies are injected into microservices' pods using ChaosMesh\footnote{https://github.com/chaos-mesh/chaos-mesh}. There are a total of 9 types of anomalies injected, including CPU stress and memory stress (detailed in Table \ref{tab:faulttype}). Leveraging this setup, we have generated a dataset comprising 90 incidents. For further details on microservices architecture and multimodal data, kindly consult the resources available at \url{https://benchmark.aiops.cn/}.

\begin{table}[H]
\centering
\caption{Fault Types}
\label{tab:faulttype}
\small
\begin{tabular}{ll} 
\toprule
\multicolumn{1}{c}{Type} & \multicolumn{1}{c}{Description}                        \\ 
\midrule
CPU Stress               & Generate threads to occupy CPU resources.         \\
Memory Stress            & Generate threads to occupy memory.                \\
Pod Failure              & Make pods inaccessible for a period of time.        \\
Network Delay            & Cause network delay for a pod.                        \\
Network Loss             & Cause packet loss in a pod's network.                 \\
Network Partition        & Network disconnection, partition.                      \\
Network Duplicate        & Cause network packet to be retransmitted.     \\
Network Corrupt          & Cause packets on network to be out of order.  \\
Network Bandwidth        & Limit the bandwidth between nodes.    \\
\bottomrule
\end{tabular}
\end{table}

\subsubsection{Evaluation Metric And Baseline Methods}

In the field of RCA, the specific location of the root cause is a critical focus for SREs. Additionally, categorizing the type of root cause is equally important, as SREs often specialize in different department like networking group or hardware group. Therefore, we have designed evaluation metrics focusing on both root cause location and fault type. Following the principle from mABC \citep{zhang2024mabc}, we consider redundant causes to be less detrimental than missing causes. Hence, we utilize two metrics: Root Cause Location Accuracy (LA) and Root Cause Type Accuracy (TA). 

\begin{equation}
{\fontsize{2}{3}\selectfont 
    LA = \frac{L_c - \sigma \times L_i}{L_t},\  TA = \frac{T_c - \sigma * T_i}{T_t}
}
\label{eq:einstein}
\end{equation}

$L_c$ and $T_c$ represent all correctly identified root cause locations and types, while $L_i$ and $T_i$ denote the incorrectly identified locations and types. $L_t$ and $T_t$ represent total number of locations and types. $\sigma$ serves as a hyperparameter with a default value of 0.1. To prevent an excessive number of root causes, we limit the maximum number of root causes to three in LLM-based methods. In addition, we employed the Average Path Length (APL) to evaluate the efficiency of the LLM Agents. APL is defined as $\frac{\sum^N_{k=1} L_k}{N}$, where $L_k$ represents the diagnosis path length of the k-th sample, and $N$ denotes the number of samples for which diagnosis was completed within the specified maximum path length.

Regarding baseline methods, we have chosen several open-source Kubernetes RCA tools, such as K8SGPT \citep{k8sgpt} and HolmesGPT \citep{holmesgpt}. Since the implementation of RCA agents is highly specific to the scenarios, they are not open-source and are challenging to migrate. Therefore, we have developed some general-purpose open-source frameworks, such as CoT \citep{wei2022chain}, ReAct \citep{yao2022react}, and Reflexion \citep{shinn2024reflexion}, to serve as our baselines.

\subsection{RQ1: Overall Performance}

\begin{table*}
\centering
\caption{Performance of different models. The best scores for each evaluation metric are bolded, and the second-best scores are underlined. Exclusive utilization of the APL metric is restricted to methodologies leveraging LLM agents. The fixed and specific accuracy of K8SGPT and HolmesGPT, i.e. 11.11, is due to their ability to handle only one type of fault.}
\label{tab:overallperformance}
\small
\begin{tabular}{>{\centering\arraybackslash}p{3.5cm} >{\centering\arraybackslash}p{2.5cm} >{\centering\arraybackslash}p{1.8cm} >{\centering\arraybackslash}p{1.8cm} >{\centering\arraybackslash}p{1.8cm} >{\centering\arraybackslash}p{1.8cm}} 
\toprule
\textbf{Model}     & \textbf{Base}          & \textbf{LA}    & \textbf{TA}    & \textbf{Average} & \textbf{APL}  \\ 
\midrule
\textbf{K8SGPT}   & GPT-3.5-Turbo & 11.11 & 11.11 & 11.11   &    -  \\
\textbf{HolmesGPT} & GPT-3.5-Turbo & 11.11 & 11.11 & 11.11   &    -  \\
\textbf{CoT}       & GPT-3.5-Turbo & 20.89 & 15.56 & 18.26   &    -  \\
\textbf{CoT}       & GPT-4-Turbo   & 36.00 & 29.22 & 32.61   &    - \\
\textbf{ReAct}     & GPT-3.5-Turbo & 13.11 & 25.22 & 19.17   &  \textbf{9.41}   \\
\textbf{ReAct}     & GPT-4-Turbo   & 47.67 & 23.33 & 35.50   &   \underline{10.76}   \\
\textbf{Reflexion} & GPT-3.5-Turbo & 21.56 & 22.22 & 21.89   &   22.38   \\
\textbf{Reflexion} & GPT-4-Turbo   & 33.67 & 24.44 & 29.06   &   28.09   \\
\textbf{Flow-of-Action}     & GPT-3.5-Turbo & \underline{54.22} & \underline{53.89} & \underline{54.06}   &   18.83   \\
\textbf{Flow-of-Action}     & GPT-4-Turbo   & \textbf{70.89} & \textbf{57.12} & \textbf{64.01}   &   15.10   \\
\bottomrule
\end{tabular}
\end{table*}

Based on Table \ref{tab:overallperformance}, our Flow-of-Action surpasses the SOTA by 23\% in the LA metric and 28\% in the TA metric. Despite the support of LLMs, K8SGPT and HolmesGPT continue to exhibit poor performance. This can be attributed to the significant limitations in the information they access. For instance, K8SGPT primarily queries Kubernetes metadata for attribute information, which is often insufficient for RCA, as faults may not necessarily manifest in metadata. CoT performs reasonably well in some common simple tasks due to the robust reasoning capabilities of LLMs. However, in RCA, where tasks are complex and diverse scenarios arise, even seasoned SREs struggle to promptly determine a series of pinpointing steps. Consequently, CoT fares poorly in the RCA domain. While ReAct integrates reasoning for each observation, the array of tools and diverse observations present challenges in rational orchestration. This is why we introduce the action set and SOP flow. Reflexion builds upon ReAct by introducing a path reflection mechanism. However, given that previous paths are predominantly incorrect, reflecting on a wealth of erroneous knowledge makes it arduous to arrive at accurate insights.

In terms of the APL metric, ReAct often erroneously identifies root causes due to a lack of proper judgment criteria, resulting in a relatively low APL. In contrast, Reflexion necessitates continuous path reflection, leading to numerous iterations and a higher APL. Flow-of-Action maintains an APL within an acceptable range, crucial for optimal performance in RCA tasks. In RCA tasks, the APL's magnitude is not fixed. Excessive values can escalate resource consumption and induce knowledge clutter, while inadequate values may lead to incomplete knowledge.

\subsection{RQ2: Impact of Action Set Size}


\begin{figure}[htbp]
\centerline{\includegraphics[scale=0.135]{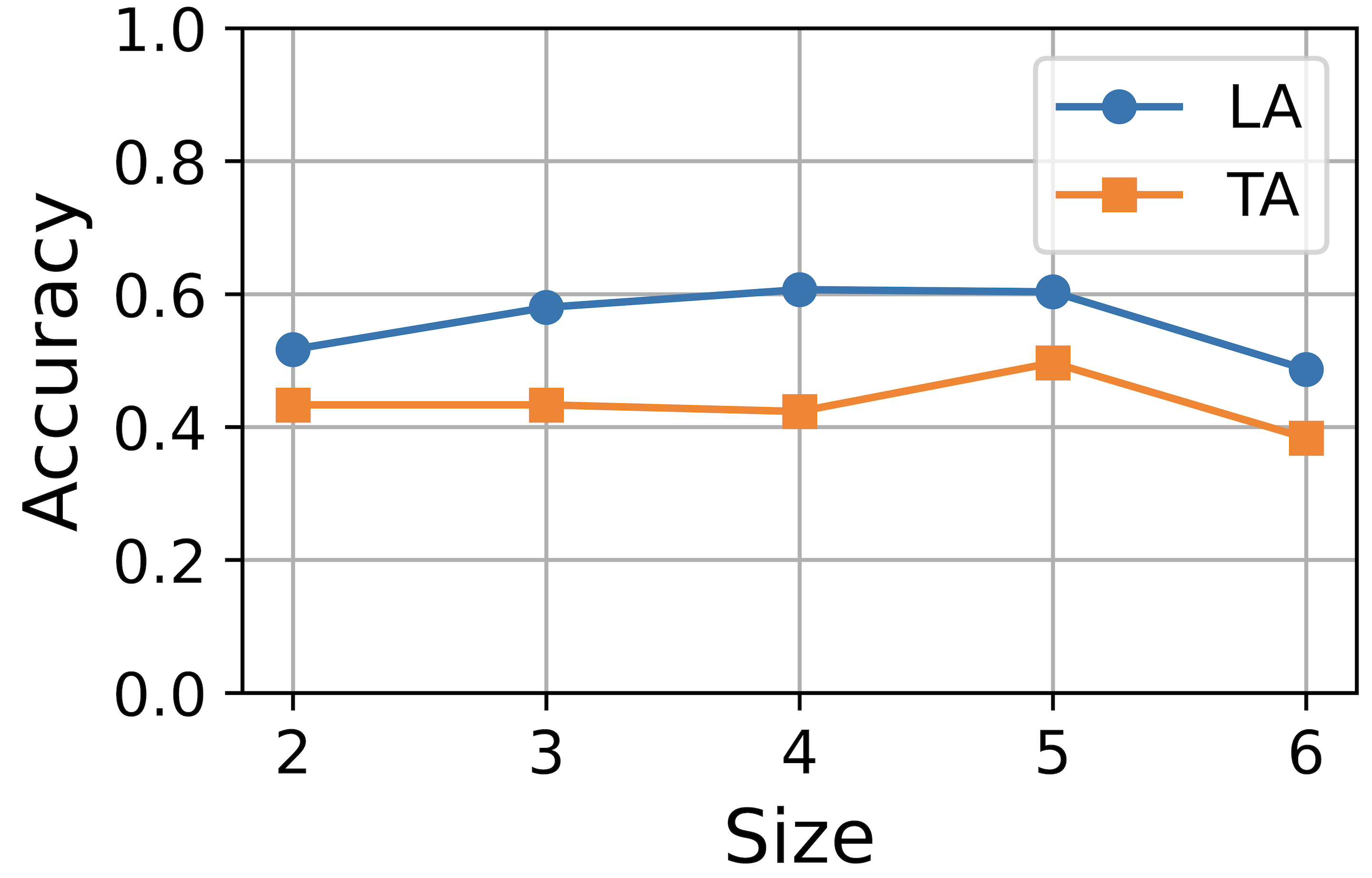}}
\caption{Accuracy of different action set sizes.}
\label{actionsetsize}
\end{figure}

As shown in Figure ~\ref{fig:model}, we have introduced the action set mechanism, where the size of the action set impacts the subsequent selection of actions. We conducted validation on a subset of the dataset and the results are shown in Figure \ref{actionsetsize}. We observed that the LA and TA remain relatively stable with changes in the action set size. This stability is attributed to the fact that, despite variations in the action set size, relevant flow tools are encompassed within the action set due to the constraints of the rules in SOP flow. Furthermore, the entire RCA process typically follows the flow, thereby minimizing significant fluctuations in accuracy. However, as the size increases, accuracy initially rises and then declines. This phenomenon occurs because smaller action sets restrict randomness, rendering the model incapable of handling complex scenarios. Conversely, larger sizes introduce more randomness, leading to a loss of control by the model. Hence, we opt for a moderately sized default value of 5 as it strikes a balance between these extremes.

\subsection{RQ3: Ablation Study}

\begin{table*}
\centering
\caption{Ablation study. The LLM backbone we use is GPT-3.5-Turbo.}
\label{tab:ablation}
\small
\begin{tabular}{>{\centering\arraybackslash}p{3.2cm} >{\centering\arraybackslash}p{1.8cm} >{\centering\arraybackslash}p{1.8cm} >{\centering\arraybackslash}p{1.8cm} >{\centering\arraybackslash}p{1.8cm}} 
\toprule
\textbf{Method}         & \textbf{LA}    & \textbf{TA}    & \textbf{Average} & \textbf{APL}    \\ 
\midrule
\textbf{Flow-of-Action}          & \textbf{54.22} & \textbf{53.89} & \textbf{54.06}   & 18.83  \\
\midrule
w/o SOP Knowledge        & 8.56  & 22.11 & 15.39   & 20.00  \\
w/o SOP Flow   & 15.11 & 39.89 & 27.50    & 19.78  \\
w/o Action Set  & 44.67 & 40.00 & 42.34   & \textbf{11.48}  \\ 
\hline
w/o ActionAgent & 32.78 & 34.56 & 33.67   & 18.42  \\
w/o ObAgent     & 40.11 & 28.67 & 34.39   & 19.31  \\
w/o JudgeAgent  & 36.11 & 33.89 & 35.00   & 20.00  \\
\bottomrule
\end{tabular}
\end{table*}

We conducted a detailed ablation study by removing each module and each agent of Flow-of-Action, with the results summarized in Table \ref{tab:ablation}. When the SOP was removed, lacking domain-specific guidance, the model relied solely on its own orchestration, essentially reverting to ReAct. The significantly low accuracy underscores the crucial role of SOP. It is worth mentioning that when SOP knowledge is removed, the SOP flow becomes ineffective as well, thus removing SOP knowledge is equivalent to removing both SOP knowledge and SOP flow.

Upon removing the prompts related to the SOP flow, we noticed a significant decrease in LA, while TA remained relatively effective. This is because SOP knowledge and relevant tools were still present and could provide type information through tools like $match\_observation$ or $match\_sop$. However, the absence of the flow hindered the complete execution of the SOP, leading to the incapacity to discern location information.

The absence of the action set rendered the model unable to make correct judgments in complex and rare scenarios. However, in most cases, the model still performed adequately, resulting in a moderate decrease in effectiveness. Without the action set, the model tended to rely more on tools determined by the flow, reducing the likelihood of excessive tool invocations and thus significantly lowering APL.

At the multi-agent level, the removal of any single agent led to a certain degree of decrease in accuracy. This is attributed to the complexity of the RCA task, where having a single agent handle all processes may lead to oversight and hallucinations. In contrast, a MAS with one main agent and multiple auxiliary agents effectively addresses this issue. The main agent can make decisions by considering the opinions of others, reducing the cognitive load and consequently achieving higher accuracy.

Regarding APL, apart from the significant impact of removing the action set, the effects of other ablations were relatively similar. This is due to the imposed limit of 20 steps to prevent unbounded loops that could render the RCA process unending.




%% file: conclusion.tex
The occurrence of frequent incidents necessitates RCA for swift issue resolution. Applying LLM agents in RCA presents numerous challenges. To address the challenges, we propose Flow-of-Action, a novel SOP-enhanced MAS. Flow-of-Action effectively leverages SOP by designing the SOP flow to alleviate hallucinations in the orchestration process. The action set mechanism efficiently tackles the challenge of selecting actions in the face of diverse observations. By employing a main agent supported by multiple auxiliary agents, Flow-of-Action further refines the delineation of responsibilities among agents, thereby enhancing the overall accuracy. Experimental results demonstrate the efficacy of Flow-of-Action.

%% file: acknowledgement.tex
This work was supported by the Chinese Academy of Sciences (241711KYSB20200023), the National Natural Science Foundation of China (62202445), and the National Natural Science Foundation of China-Research Grants Council (RGC) Joint Research Scheme (62321166652).

%% file: main.bbl

\begin{thebibliography}{37}


\ifx \showCODEN    \undefined \def \showCODEN     #1{\unskip}     \fi
\ifx \showDOI      \undefined \def \showDOI       #1{#1}\fi
\ifx \showISBNx    \undefined \def \showISBNx     #1{\unskip}     \fi
\ifx \showISBNxiii \undefined \def \showISBNxiii  #1{\unskip}     \fi
\ifx \showISSN     \undefined \def \showISSN      #1{\unskip}     \fi
\ifx \showLCCN     \undefined \def \showLCCN      #1{\unskip}     \fi
\ifx \shownote     \undefined \def \shownote      #1{#1}          \fi
\ifx \showarticletitle \undefined \def \showarticletitle #1{#1}   \fi
\ifx \showURL      \undefined \def \showURL       {\relax}        \fi
\providecommand\bibfield[2]{#2}
\providecommand\bibinfo[2]{#2}
\providecommand\natexlab[1]{#1}
\providecommand\showeprint[2][]{arXiv:#2}

\bibitem[Chakraborty et~al\mbox{.}(2023)]%
        {micro5}
\bibfield{author}{\bibinfo{person}{Sarthak Chakraborty}, \bibinfo{person}{Shaddy Garg}, \bibinfo{person}{Shubham Agarwal}, \bibinfo{person}{Ayush Chauhan}, {and} \bibinfo{person}{Shiv~Kumar Saini}.} \bibinfo{year}{2023}\natexlab{}.
\newblock \showarticletitle{Causil: Causal graph for instance level microservice data}. In \bibinfo{booktitle}{\emph{Proceedings of the ACM Web Conference 2023}}. \bibinfo{pages}{2905--2915}.
\newblock


\bibitem[Chen et~al\mbox{.}(2024a)]%
        {chen2024microfi}
\bibfield{author}{\bibinfo{person}{Hongyang Chen}, \bibinfo{person}{Pengfei Chen}, \bibinfo{person}{Guangba Yu}, \bibinfo{person}{Xiaoyun Li}, \bibinfo{person}{Zilong He}, {and} \bibinfo{person}{Huxing Zhang}.} \bibinfo{year}{2024}\natexlab{a}.
\newblock \showarticletitle{MicroFI: Non-Intrusive and Prioritized Request-Level Fault Injection for Microservice Applications}.
\newblock \bibinfo{journal}{\emph{IEEE Transactions on Dependable and Secure Computing}} (\bibinfo{year}{2024}).
\newblock


\bibitem[Chen et~al\mbox{.}(2024b)]%
        {chen2024automatic}
\bibfield{author}{\bibinfo{person}{Yinfang Chen}, \bibinfo{person}{Huaibing Xie}, \bibinfo{person}{Minghua Ma}, \bibinfo{person}{Yu Kang}, \bibinfo{person}{Xin Gao}, \bibinfo{person}{Liu Shi}, \bibinfo{person}{Yunjie Cao}, \bibinfo{person}{Xuedong Gao}, \bibinfo{person}{Hao Fan}, \bibinfo{person}{Ming Wen}, {et~al\mbox{.}}} \bibinfo{year}{2024}\natexlab{b}.
\newblock \showarticletitle{Automatic root cause analysis via large language models for cloud incidents}. In \bibinfo{booktitle}{\emph{Proceedings of the Nineteenth European Conference on Computer Systems}}. \bibinfo{pages}{674--688}.
\newblock


\bibitem[Ding et~al\mbox{.}(2023)]%
        {rca3}
\bibfield{author}{\bibinfo{person}{Ruomeng Ding}, \bibinfo{person}{Chaoyun Zhang}, \bibinfo{person}{Lu Wang}, \bibinfo{person}{Yong Xu}, \bibinfo{person}{Minghua Ma}, \bibinfo{person}{Xiaomin Wu}, \bibinfo{person}{Meng Zhang}, \bibinfo{person}{Qingjun Chen}, \bibinfo{person}{Xin Gao}, \bibinfo{person}{Xuedong Gao}, {et~al\mbox{.}}} \bibinfo{year}{2023}\natexlab{}.
\newblock \showarticletitle{TraceDiag: Adaptive, Interpretable, and Efficient Root Cause Analysis on Large-Scale Microservice Systems}. In \bibinfo{booktitle}{\emph{Proceedings of the 31st ACM Joint European Software Engineering Conference and Symposium on the Foundations of Software Engineering}}. \bibinfo{pages}{1762--1773}.
\newblock


\bibitem[Hong et~al\mbox{.}(2023)]%
        {hong2023metagpt}
\bibfield{author}{\bibinfo{person}{Sirui Hong}, \bibinfo{person}{Xiawu Zheng}, \bibinfo{person}{Jonathan Chen}, \bibinfo{person}{Yuheng Cheng}, \bibinfo{person}{Jinlin Wang}, \bibinfo{person}{Ceyao Zhang}, \bibinfo{person}{Zili Wang}, \bibinfo{person}{Steven Ka~Shing Yau}, \bibinfo{person}{Zijuan Lin}, \bibinfo{person}{Liyang Zhou}, {et~al\mbox{.}}} \bibinfo{year}{2023}\natexlab{}.
\newblock \showarticletitle{Metagpt: Meta programming for multi-agent collaborative framework}.
\newblock \bibinfo{journal}{\emph{arXiv preprint arXiv:2308.00352}} (\bibinfo{year}{2023}).
\newblock


\bibitem[Ikram et~al\mbox{.}(2022)]%
        {rca1}
\bibfield{author}{\bibinfo{person}{Azam Ikram}, \bibinfo{person}{Sarthak Chakraborty}, \bibinfo{person}{Subrata Mitra}, \bibinfo{person}{Shiv Saini}, \bibinfo{person}{Saurabh Bagchi}, {and} \bibinfo{person}{Murat Kocaoglu}.} \bibinfo{year}{2022}\natexlab{}.
\newblock \showarticletitle{Root cause analysis of failures in microservices through causal discovery}.
\newblock \bibinfo{journal}{\emph{Advances in Neural Information Processing Systems}}  \bibinfo{volume}{35} (\bibinfo{year}{2022}), \bibinfo{pages}{31158--31170}.
\newblock


\bibitem[Jeong et~al\mbox{.}(2024)]%
        {jeong2024adaptive}
\bibfield{author}{\bibinfo{person}{Soyeong Jeong}, \bibinfo{person}{Jinheon Baek}, \bibinfo{person}{Sukmin Cho}, \bibinfo{person}{Sung~Ju Hwang}, {and} \bibinfo{person}{Jong~C Park}.} \bibinfo{year}{2024}\natexlab{}.
\newblock \showarticletitle{Adaptive-rag: Learning to adapt retrieval-augmented large language models through question complexity}.
\newblock \bibinfo{journal}{\emph{arXiv preprint arXiv:2403.14403}} (\bibinfo{year}{2024}).
\newblock


\bibitem[Jiang et~al\mbox{.}(2023)]%
        {micro4}
\bibfield{author}{\bibinfo{person}{Xinrui Jiang}, \bibinfo{person}{Yicheng Pan}, \bibinfo{person}{Meng Ma}, {and} \bibinfo{person}{Ping Wang}.} \bibinfo{year}{2023}\natexlab{}.
\newblock \showarticletitle{Look Deep into the Microservice System Anomaly through Very Sparse Logs}. In \bibinfo{booktitle}{\emph{Proceedings of the ACM Web Conference 2023}}. \bibinfo{pages}{2970--2978}.
\newblock


\bibitem[k8sgpt ai(2023)]%
        {k8sgpt}
\bibfield{author}{\bibinfo{person}{k8sgpt ai}.} \bibinfo{year}{2023}\natexlab{}.
\newblock \bibinfo{title}{k8sgpt}.
\newblock \bibinfo{howpublished}{\url{https://github.com/k8sgpt-ai/k8sgpt}}.
\newblock


\bibitem[Labring(2023)]%
        {fastgpt}
\bibfield{author}{\bibinfo{person}{Labring}.} \bibinfo{year}{2023}\natexlab{}.
\newblock \bibinfo{title}{FastGPT}.
\newblock \bibinfo{howpublished}{\url{https://github.com/labring/FastGPT}}.
\newblock


\bibitem[Li et~al\mbox{.}(2023)]%
        {li2023chainofcode}
\bibfield{author}{\bibinfo{person}{Chengshu Li}, \bibinfo{person}{Jacky Liang}, \bibinfo{person}{Andy Zeng}, \bibinfo{person}{Xinyun Chen}, \bibinfo{person}{Karol Hausman}, \bibinfo{person}{Dorsa Sadigh}, \bibinfo{person}{Sergey Levine}, \bibinfo{person}{Li Fei-Fei}, \bibinfo{person}{Fei Xia}, {and} \bibinfo{person}{Brian Ichter}.} \bibinfo{year}{2023}\natexlab{}.
\newblock \showarticletitle{Chain of code: Reasoning with a language model-augmented code emulator}.
\newblock \bibinfo{journal}{\emph{arXiv preprint arXiv:2312.04474}} (\bibinfo{year}{2023}).
\newblock


\bibitem[Li et~al\mbox{.}(2022)]%
        {li2022dejavu}
\bibfield{author}{\bibinfo{person}{Zeyan Li}, \bibinfo{person}{Nengwen Zhao}, \bibinfo{person}{Mingjie Li}, \bibinfo{person}{Xianglin Lu}, \bibinfo{person}{Lixin Wang}, \bibinfo{person}{Dongdong Chang}, \bibinfo{person}{Xiaohui Nie}, \bibinfo{person}{Li Cao}, \bibinfo{person}{Wenchi Zhang}, \bibinfo{person}{Kaixin Sui}, {et~al\mbox{.}}} \bibinfo{year}{2022}\natexlab{}.
\newblock \showarticletitle{Actionable and interpretable fault localization for recurring failures in online service systems}. In \bibinfo{booktitle}{\emph{Proceedings of the 30th ACM Joint European Software Engineering Conference and Symposium on the Foundations of Software Engineering}}. \bibinfo{pages}{996--1008}.
\newblock


\bibitem[Lin et~al\mbox{.}(2024)]%
        {rca4}
\bibfield{author}{\bibinfo{person}{Cheng-Ming Lin}, \bibinfo{person}{Ching Chang}, \bibinfo{person}{Wei-Yao Wang}, \bibinfo{person}{Kuang-Da Wang}, {and} \bibinfo{person}{Wen-Chih Peng}.} \bibinfo{year}{2024}\natexlab{}.
\newblock \showarticletitle{Root Cause Analysis in Microservice Using Neural Granger Causal Discovery}. In \bibinfo{booktitle}{\emph{Proceedings of the AAAI Conference on Artificial Intelligence}}, Vol.~\bibinfo{volume}{38}. \bibinfo{pages}{206--213}.
\newblock


\bibitem[Misiakos et~al\mbox{.}(2024)]%
        {metric6}
\bibfield{author}{\bibinfo{person}{Panagiotis Misiakos}, \bibinfo{person}{Chris Wendler}, {and} \bibinfo{person}{Markus P{\"u}schel}.} \bibinfo{year}{2024}\natexlab{}.
\newblock \showarticletitle{Learning DAGs from data with few root causes}.
\newblock \bibinfo{journal}{\emph{Advances in Neural Information Processing Systems}}  \bibinfo{volume}{36} (\bibinfo{year}{2024}).
\newblock


\bibitem[Pan et~al\mbox{.}(2023)]%
        {pan2023logiclm}
\bibfield{author}{\bibinfo{person}{Liangming Pan}, \bibinfo{person}{Alon Albalak}, \bibinfo{person}{Xinyi Wang}, {and} \bibinfo{person}{William~Yang Wang}.} \bibinfo{year}{2023}\natexlab{}.
\newblock \showarticletitle{Logic-lm: Empowering large language models with symbolic solvers for faithful logical reasoning}.
\newblock \bibinfo{journal}{\emph{arXiv preprint arXiv:2305.12295}} (\bibinfo{year}{2023}).
\newblock


\bibitem[Qin et~al\mbox{.}(2023)]%
        {qin2023toolllm}
\bibfield{author}{\bibinfo{person}{Yujia Qin}, \bibinfo{person}{Shihao Liang}, \bibinfo{person}{Yining Ye}, \bibinfo{person}{Kunlun Zhu}, \bibinfo{person}{Lan Yan}, \bibinfo{person}{Yaxi Lu}, \bibinfo{person}{Yankai Lin}, \bibinfo{person}{Xin Cong}, \bibinfo{person}{Xiangru Tang}, \bibinfo{person}{Bill Qian}, {et~al\mbox{.}}} \bibinfo{year}{2023}\natexlab{}.
\newblock \showarticletitle{Toolllm: Facilitating large language models to master 16000+ real-world apis}.
\newblock \bibinfo{journal}{\emph{arXiv preprint arXiv:2307.16789}} (\bibinfo{year}{2023}).
\newblock


\bibitem[Radford(2018)]%
        {radford2018gpt}
\bibfield{author}{\bibinfo{person}{Alec Radford}.} \bibinfo{year}{2018}\natexlab{}.
\newblock \showarticletitle{Improving language understanding by generative pre-training}.
\newblock  (\bibinfo{year}{2018}).
\newblock


\bibitem[Radford et~al\mbox{.}(2019)]%
        {radford2019gpt2}
\bibfield{author}{\bibinfo{person}{Alec Radford}, \bibinfo{person}{Jeffrey Wu}, \bibinfo{person}{Rewon Child}, \bibinfo{person}{David Luan}, \bibinfo{person}{Dario Amodei}, \bibinfo{person}{Ilya Sutskever}, {et~al\mbox{.}}} \bibinfo{year}{2019}\natexlab{}.
\newblock \showarticletitle{Language models are unsupervised multitask learners}.
\newblock \bibinfo{journal}{\emph{OpenAI blog}} \bibinfo{volume}{1}, \bibinfo{number}{8} (\bibinfo{year}{2019}), \bibinfo{pages}{9}.
\newblock


\bibitem[robusta dev(2024)]%
        {holmesgpt}
\bibfield{author}{\bibinfo{person}{robusta dev}.} \bibinfo{year}{2024}\natexlab{}.
\newblock \bibinfo{title}{holmesgpt}.
\newblock \bibinfo{howpublished}{\url{https://github.com/robusta-dev/holmesgpt}}.
\newblock


\bibitem[Rosenberg and Moonen(2020)]%
        {log2}
\bibfield{author}{\bibinfo{person}{Carl~Martin Rosenberg} {and} \bibinfo{person}{Leon Moonen}.} \bibinfo{year}{2020}\natexlab{}.
\newblock \showarticletitle{Spectrum-based log diagnosis}. In \bibinfo{booktitle}{\emph{Proceedings of the 14th ACM/IEEE International Symposium on Empirical Software Engineering and Measurement (ESEM)}}. \bibinfo{pages}{1--12}.
\newblock


\bibitem[Schick et~al\mbox{.}(2024)]%
        {schick2024toolformer}
\bibfield{author}{\bibinfo{person}{Timo Schick}, \bibinfo{person}{Jane Dwivedi-Yu}, \bibinfo{person}{Roberto Dess{\`\i}}, \bibinfo{person}{Roberta Raileanu}, \bibinfo{person}{Maria Lomeli}, \bibinfo{person}{Eric Hambro}, \bibinfo{person}{Luke Zettlemoyer}, \bibinfo{person}{Nicola Cancedda}, {and} \bibinfo{person}{Thomas Scialom}.} \bibinfo{year}{2024}\natexlab{}.
\newblock \showarticletitle{Toolformer: Language models can teach themselves to use tools}.
\newblock \bibinfo{journal}{\emph{Advances in Neural Information Processing Systems}}  \bibinfo{volume}{36} (\bibinfo{year}{2024}).
\newblock


\bibitem[Shen et~al\mbox{.}(2023)]%
        {deepflow}
\bibfield{author}{\bibinfo{person}{Junxian Shen}, \bibinfo{person}{Han Zhang}, \bibinfo{person}{Yang Xiang}, \bibinfo{person}{Xingang Shi}, \bibinfo{person}{Xinrui Li}, \bibinfo{person}{Yunxi Shen}, \bibinfo{person}{Zijian Zhang}, \bibinfo{person}{Yongxiang Wu}, \bibinfo{person}{Xia Yin}, \bibinfo{person}{Jilong Wang}, {et~al\mbox{.}}} \bibinfo{year}{2023}\natexlab{}.
\newblock \showarticletitle{Network-centric distributed tracing with deepflow: Troubleshooting your microservices in zero code}. In \bibinfo{booktitle}{\emph{Proceedings of the ACM SIGCOMM 2023 Conference}}. \bibinfo{pages}{420--437}.
\newblock


\bibitem[Shinn et~al\mbox{.}(2024)]%
        {shinn2024reflexion}
\bibfield{author}{\bibinfo{person}{Noah Shinn}, \bibinfo{person}{Federico Cassano}, \bibinfo{person}{Ashwin Gopinath}, \bibinfo{person}{Karthik Narasimhan}, {and} \bibinfo{person}{Shunyu Yao}.} \bibinfo{year}{2024}\natexlab{}.
\newblock \showarticletitle{Reflexion: Language agents with verbal reinforcement learning}.
\newblock \bibinfo{journal}{\emph{Advances in Neural Information Processing Systems}}  \bibinfo{volume}{36} (\bibinfo{year}{2024}).
\newblock


\bibitem[Somashekar et~al\mbox{.}(2024)]%
        {micro3}
\bibfield{author}{\bibinfo{person}{Gagan Somashekar}, \bibinfo{person}{Anurag Dutt}, \bibinfo{person}{Mainak Adak}, \bibinfo{person}{Tania Lorido~Botran}, {and} \bibinfo{person}{Anshul Gandhi}.} \bibinfo{year}{2024}\natexlab{}.
\newblock \showarticletitle{GAMMA: Graph Neural Network-Based Multi-Bottleneck Localization for Microservices Applications}. In \bibinfo{booktitle}{\emph{Proceedings of the ACM on Web Conference 2024}}. \bibinfo{pages}{3085--3095}.
\newblock


\bibitem[Tuli et~al\mbox{.}(2022)]%
        {tuli2022tranad}
\bibfield{author}{\bibinfo{person}{Shreshth Tuli}, \bibinfo{person}{Giuliano Casale}, {and} \bibinfo{person}{Nicholas~R Jennings}.} \bibinfo{year}{2022}\natexlab{}.
\newblock \showarticletitle{Tranad: Deep transformer networks for anomaly detection in multivariate time series data}.
\newblock \bibinfo{journal}{\emph{arXiv preprint arXiv:2201.07284}} (\bibinfo{year}{2022}).
\newblock


\bibitem[Wang et~al\mbox{.}(2023)]%
        {micro1}
\bibfield{author}{\bibinfo{person}{Lu Wang}, \bibinfo{person}{Chaoyun Zhang}, \bibinfo{person}{Ruomeng Ding}, \bibinfo{person}{Yong Xu}, \bibinfo{person}{Qihang Chen}, \bibinfo{person}{Wentao Zou}, \bibinfo{person}{Qingjun Chen}, \bibinfo{person}{Meng Zhang}, \bibinfo{person}{Xuedong Gao}, \bibinfo{person}{Hao Fan}, {et~al\mbox{.}}} \bibinfo{year}{2023}\natexlab{}.
\newblock \showarticletitle{Root cause analysis for microservice systems via hierarchical reinforcement learning from human feedback}. In \bibinfo{booktitle}{\emph{Proceedings of the 29th ACM SIGKDD Conference on Knowledge Discovery and Data Mining}}. \bibinfo{pages}{5116--5125}.
\newblock


\bibitem[Wang et~al\mbox{.}(2024a)]%
        {wang2024large}
\bibfield{author}{\bibinfo{person}{Zexin Wang}, \bibinfo{person}{Jianhui Li}, \bibinfo{person}{Minghua Ma}, \bibinfo{person}{Ze Li}, \bibinfo{person}{Yu Kang}, \bibinfo{person}{Chaoyun Zhang}, \bibinfo{person}{Chetan Bansal}, \bibinfo{person}{Murali Chintalapati}, \bibinfo{person}{Saravan Rajmohan}, \bibinfo{person}{Qingwei Lin}, {et~al\mbox{.}}} \bibinfo{year}{2024}\natexlab{a}.
\newblock \showarticletitle{Large Language Models Can Provide Accurate and Interpretable Incident Triage}. In \bibinfo{booktitle}{\emph{2024 IEEE 35th International Symposium on Software Reliability Engineering (ISSRE)}}. IEEE, \bibinfo{pages}{523--534}.
\newblock


\bibitem[Wang et~al\mbox{.}(2024b)]%
        {wang2024revisiting}
\bibfield{author}{\bibinfo{person}{Zexin Wang}, \bibinfo{person}{Changhua Pei}, \bibinfo{person}{Minghua Ma}, \bibinfo{person}{Xin Wang}, \bibinfo{person}{Zhihan Li}, \bibinfo{person}{Dan Pei}, \bibinfo{person}{Saravan Rajmohan}, \bibinfo{person}{Dongmei Zhang}, \bibinfo{person}{Qingwei Lin}, \bibinfo{person}{Haiming Zhang}, {et~al\mbox{.}}} \bibinfo{year}{2024}\natexlab{b}.
\newblock \showarticletitle{Revisiting VAE for Unsupervised Time Series Anomaly Detection: A Frequency Perspective}. In \bibinfo{booktitle}{\emph{Proceedings of the ACM on Web Conference 2024}}. \bibinfo{pages}{3096--3105}.
\newblock


\bibitem[Wei et~al\mbox{.}(2022)]%
        {wei2022chain}
\bibfield{author}{\bibinfo{person}{Jason Wei}, \bibinfo{person}{Xuezhi Wang}, \bibinfo{person}{Dale Schuurmans}, \bibinfo{person}{Maarten Bosma}, \bibinfo{person}{Fei Xia}, \bibinfo{person}{Ed Chi}, \bibinfo{person}{Quoc~V Le}, \bibinfo{person}{Denny Zhou}, {et~al\mbox{.}}} \bibinfo{year}{2022}\natexlab{}.
\newblock \showarticletitle{Chain-of-thought prompting elicits reasoning in large language models}.
\newblock \bibinfo{journal}{\emph{Advances in neural information processing systems}}  \bibinfo{volume}{35} (\bibinfo{year}{2022}), \bibinfo{pages}{24824--24837}.
\newblock


\bibitem[Wu et~al\mbox{.}(2020)]%
        {rca2}
\bibfield{author}{\bibinfo{person}{Li Wu}, \bibinfo{person}{Johan Tordsson}, \bibinfo{person}{Erik Elmroth}, {and} \bibinfo{person}{Odej Kao}.} \bibinfo{year}{2020}\natexlab{}.
\newblock \showarticletitle{Microrca: Root cause localization of performance issues in microservices}. In \bibinfo{booktitle}{\emph{NOMS 2020-2020 IEEE/IFIP Network Operations and Management Symposium}}. IEEE, \bibinfo{pages}{1--9}.
\newblock


\bibitem[Xu et~al\mbox{.}(2018)]%
        {donut}
\bibfield{author}{\bibinfo{person}{Haowen Xu}, \bibinfo{person}{Wenxiao Chen}, \bibinfo{person}{Nengwen Zhao}, \bibinfo{person}{Zeyan Li}, \bibinfo{person}{Jiahao Bu}, \bibinfo{person}{Zhihan Li}, \bibinfo{person}{Ying Liu}, \bibinfo{person}{Youjian Zhao}, \bibinfo{person}{Dan Pei}, \bibinfo{person}{Yang Feng}, {et~al\mbox{.}}} \bibinfo{year}{2018}\natexlab{}.
\newblock \showarticletitle{Unsupervised anomaly detection via variational auto-encoder for seasonal kpis in web applications}. In \bibinfo{booktitle}{\emph{Proceedings of the 2018 world wide web conference}}. \bibinfo{pages}{187--196}.
\newblock


\bibitem[Yao et~al\mbox{.}(2022)]%
        {yao2022react}
\bibfield{author}{\bibinfo{person}{Shunyu Yao}, \bibinfo{person}{Jeffrey Zhao}, \bibinfo{person}{Dian Yu}, \bibinfo{person}{Nan Du}, \bibinfo{person}{Izhak Shafran}, \bibinfo{person}{Karthik Narasimhan}, {and} \bibinfo{person}{Yuan Cao}.} \bibinfo{year}{2022}\natexlab{}.
\newblock \showarticletitle{React: Synergizing reasoning and acting in language models}.
\newblock \bibinfo{journal}{\emph{arXiv preprint arXiv:2210.03629}} (\bibinfo{year}{2022}).
\newblock


\bibitem[Yao et~al\mbox{.}(2024a)]%
        {yao2024chain}
\bibfield{author}{\bibinfo{person}{Zhenhe Yao}, \bibinfo{person}{Changhua Pei}, \bibinfo{person}{Wenxiao Chen}, \bibinfo{person}{Hanzhang Wang}, \bibinfo{person}{Liangfei Su}, \bibinfo{person}{Huai Jiang}, \bibinfo{person}{Zhe Xie}, \bibinfo{person}{Xiaohui Nie}, {and} \bibinfo{person}{Dan Pei}.} \bibinfo{year}{2024}\natexlab{a}.
\newblock \showarticletitle{Chain-of-Event: Interpretable Root Cause Analysis for Microservices through Automatically Learning Weighted Event Causal Graph}. In \bibinfo{booktitle}{\emph{Companion Proceedings of the 32nd ACM International Conference on the Foundations of Software Engineering}}. \bibinfo{pages}{50--61}.
\newblock


\bibitem[Yao et~al\mbox{.}(2024b)]%
        {yaosparserca}
\bibfield{author}{\bibinfo{person}{Zhenhe Yao}, \bibinfo{person}{Haowei Ye}, \bibinfo{person}{Changhua Pei}, \bibinfo{person}{Guang Cheng}, \bibinfo{person}{Guangpei Wang}, \bibinfo{person}{Zhiwei Liu}, \bibinfo{person}{Hongwei Chen}, \bibinfo{person}{Hang Cui}, \bibinfo{person}{Zeyan Li}, \bibinfo{person}{Jianhui Li}, {et~al\mbox{.}}} \bibinfo{year}{2024}\natexlab{b}.
\newblock \showarticletitle{SparseRCA: Unsupervised Root Cause Analysis in Sparse Microservice Testing Traces}. In \bibinfo{booktitle}{\emph{2024 IEEE 35st International Symposium on Software Reliability Engineering (ISSRE)}}.
\newblock


\bibitem[Yu et~al\mbox{.}(2023)]%
        {yu2023nezha}
\bibfield{author}{\bibinfo{person}{Guangba Yu}, \bibinfo{person}{Pengfei Chen}, \bibinfo{person}{Yufeng Li}, \bibinfo{person}{Hongyang Chen}, \bibinfo{person}{Xiaoyun Li}, {and} \bibinfo{person}{Zibin Zheng}.} \bibinfo{year}{2023}\natexlab{}.
\newblock \showarticletitle{Nezha: Interpretable fine-grained root causes analysis for microservices on multi-modal observability data}. In \bibinfo{booktitle}{\emph{Proceedings of the 31st ACM Joint European Software Engineering Conference and Symposium on the Foundations of Software Engineering}}. \bibinfo{pages}{553--565}.
\newblock


\bibitem[Zhang et~al\mbox{.}(2024)]%
        {zhang2024mabc}
\bibfield{author}{\bibinfo{person}{Wei Zhang}, \bibinfo{person}{Hongcheng Guo}, \bibinfo{person}{Jian Yang}, \bibinfo{person}{Yi Zhang}, \bibinfo{person}{Chaoran Yan}, \bibinfo{person}{Zhoujin Tian}, \bibinfo{person}{Hangyuan Ji}, \bibinfo{person}{Zhoujun Li}, \bibinfo{person}{Tongliang Li}, \bibinfo{person}{Tieqiao Zheng}, {et~al\mbox{.}}} \bibinfo{year}{2024}\natexlab{}.
\newblock \showarticletitle{mABC: multi-Agent Blockchain-Inspired Collaboration for root cause analysis in micro-services architecture}.
\newblock \bibinfo{journal}{\emph{arXiv preprint arXiv:2404.12135}} (\bibinfo{year}{2024}).
\newblock


\bibitem[Zheng et~al\mbox{.}(2024)]%
        {micro2}
\bibfield{author}{\bibinfo{person}{Lecheng Zheng}, \bibinfo{person}{Zhengzhang Chen}, \bibinfo{person}{Jingrui He}, {and} \bibinfo{person}{Haifeng Chen}.} \bibinfo{year}{2024}\natexlab{}.
\newblock \showarticletitle{MULAN: Multi-modal Causal Structure Learning and Root Cause Analysis for Microservice Systems}. In \bibinfo{booktitle}{\emph{Proceedings of the ACM on Web Conference 2024}}. \bibinfo{pages}{4107--4116}.
\newblock


\end{thebibliography}
